\documentclass[aps,twocolumn,superscriptaddress,prd]{revtex4-2}
\usepackage{amsmath}
\usepackage{amssymb}
\usepackage{latexsym}
\usepackage{amsfonts}
\usepackage{epsfig}
\usepackage{psfrag}
\usepackage{graphicx}
\usepackage{color}
\usepackage{hyperref}

\usepackage{graphicx}

\usepackage{amsmath}

\newcommand{\bea}{\begin{eqnarray}}
\newcommand{\ea}{\end{eqnarray}}
\newcommand{\eea}{\end{eqnarray}}

\begin{document}
 
 \title{Optical absorption and carrier multiplication at graphene edges in a magnetic field}

\author{Friedemann Queisser}

\affiliation{Helmholtz-Zentrum Dresden-Rossendorf, 
Bautzner Landstra{\ss}e 400, 01328 Dresden, Germany,}

\affiliation{Institut f\"ur Theoretische Physik, 
Technische Universit\"at Dresden, 01062 Dresden, Germany,}

\author{Sascha Lang}

\affiliation{Helmholtz-Zentrum Dresden-Rossendorf, 
Bautzner Landstra{\ss}e 400, 01328 Dresden, Germany,}

\affiliation{Fakult\"at f\"ur Physik, Universit\"at Duisburg-Essen, Lotharstra\ss e 1, 47057 Duisburg, Germany}

\author{Ralf Sch\"utzhold}

\affiliation{Helmholtz-Zentrum Dresden-Rossendorf, 
Bautzner Landstra{\ss}e 400, 01328 Dresden, Germany,}

\affiliation{Institut f\"ur Theoretische Physik, 
Technische Universit\"at Dresden, 01062 Dresden, Germany,}

\date{\today}

\begin{abstract}
We study optical absorption at graphene edges in a transversal magnetic field.
%
%
The magnetic field bends the trajectories of particle- and 
hole excitations into antipodal direction which generates a directed current.
We find a rather strong amplification of the edge current by
impact ionization processes.
More concretely, the primary absorption and the subsequent carrier multiplication is 
analyzed for a graphene fold and a zigzag edge.
We identify exact and approximate selection rules and discuss the dependence of the 
decay rates on the initial state.

\end{abstract}


\maketitle

\section{Introduction}

Graphene offers a large mean free path, a high charge carrier mobility \cite{BGB08,MGM11,NHP09} and a broad absorption bandwidth \cite{NBG08}
which makes it a promising candidate for optoelectronic applications
\cite{MSB19,BSHF10,KMA14}.
Photodetectors and graphene-based solar cells
require the generation of a directed photocurrent \cite{SRM11,CDT20,MLS19,SOF16,SAJ12}.
As proposed in \cite{QS13}
a directed charge current 
can be generated when photons are absorbed in the vicinity of 
a graphene edge which is subject to a perpendicular magnetic field.
%
This proposal was experimentally confirmed by measuring the optical response of a suspended graphene layer in a perpendicular magnetic field 
~\cite{SKG17}.
Surprisingly, it was found that the generated edge current adopts much larger values 
than naively expected from the photon absorption probability.
More specifically, a 7-fold larger current was observed than naively expected from the absorption probability of graphene monolayers
which is about $2.3\%$ \cite{PS08,KM16,MSW08}.
This strong enhancement was dedicated to secondary particle-hole generation via Auger scattering processes \cite{T13,P14,M15,BTM13}.
In total, we claim that this giant magneto-photoelectric current
can be understood within a two-step process.
First, photons are absorbed into the dispersive edge modes
and generate electrons-hole pairs with energies in the eV regime.
The magnetic field bends the particles and holes in antipodal directions which induces a directed current.
Subsequently, impact ionization at the graphene edge generates further charge carriers which then also
contribute to the overall current.

It is well-known that the bare photoeffect in graphene
can already be understood in the 
%
The details of this primary particle-hole generation at graphene edges will be 
discussed in section \ref{seccond} for two different 
geometries, namely a graphene fold and a zigzag edge.
The subsequent charge carrier multiplication due to 
impact ionization will be analyzed in section \ref{secparthole}.
In particular we shall give estimates for the 
charge carrier multiplication rates and discuss 
various selection rules for the decay channels.
Beforehand we briefly recapitulate the form of the 
effective Dirac equations of single layer graphene in 
section \ref{secspinors}.

\section{Dirac spinors}\label{secspinors}

The Carbon atoms of Graphene are arranged in a hexagonal structure
which can be seen as a triangular lattice with two sites in a unit cell.
The electronic wave function is localized on two sublattices which we denote 
with $A$ and $B$.
In a tight-binding model, the stationary Schr\"odinger equation can be written as \cite{A05}
\begin{align}
-\gamma_0 \sum_l \psi_B(\mathbf{r}_A-\mathbf{\tau}_l)&=E\psi_A(\mathbf{r}_A)\label{tight1}\\
-\gamma_0 \sum_l \psi_A(\mathbf{r}_B+\mathbf{\tau}_l)&=E\psi_B(\mathbf{r}_B)\label{tight2}
\end{align}
where $\gamma_0$ is the transfer integral between neighboring
carbon atoms and the vectors connecting the neighboring atom positions are given by
$\mathbf{\tau}_1=a_0(0,1/\sqrt{3})$, $\mathbf{\tau}_2=a_0(-1/2,-1/(2\sqrt{3}))$ and 
$\mathbf{\tau}_3=a_0(1/2,-1/2\sqrt{3})$ with $a_0\approx 2.46$ \r{A} being the lattice 
constant.
The coordinate system was chosen such that the graphene sheet being cut parallel to the $x$-axis has a zigzag-edge.
The energy dispersion vanishes linearly around the so-called 
Dirac points $\mathbf{K}=\frac{2\pi}{a_0}(1/3,1/\sqrt{3})$ and $\mathbf{K'}=\frac{2\pi}{a_0}(-1/3,1/\sqrt{3})$.
Since we want to describe excitations around the Fermi level $E=0$, we choose the following ansatz for the electron wavefunction
\begin{align}
\Psi_A(\mathbf{r}_A)&=e^{i\mathbf{K}\cdot \mathbf{r}_A}\psi^{K}(\mathbf{r}_A)+e^{i\mathbf{K}'\cdot \mathbf{r}_A}\psi^{K'}(\mathbf{r}_A)\label{Alattice} \\
\Psi_B(\mathbf{r}_A)&=e^{i\frac{2\pi}{3}}e^{i\mathbf{K}\cdot \mathbf{r}_B}\psi^{K}(\mathbf{r}_B)+e^{i\frac{2\pi}{3}}e^{i\mathbf{K}'\cdot \mathbf{r}_B}\psi^{K'}(\mathbf{r}_B)\,,\label{Blattice}
\end{align}
where the phases $e^{i\mathbf{K}\cdot \mathbf{r}_{A,B}}$
and $e^{i\mathbf{K'}\cdot \mathbf{r}_{A,B}}$ are highly oscillating and $\psi^{K}_{A,B}$ as well as $\psi^{K'}_{A,B}$ are slowly varying envelope functions.
The envelope functions can be grouped in two-component spinors, i.e.
$\Psi^K=[\psi^{K}_A,\psi^{K}_B]^T$ and $\Psi^{K'}=[\psi^{K'}_A,\psi^{K'}_B]^T$, which satisfy 
2-1-dimensional Dirac equations \cite{BF06},
\begin{align}\label{DiracK}
v_F\begin{pmatrix}
0 &-\hat{p}_x+i \hat{p}_y\\
-\hat{p}_x-i \hat{p}_y &0
\end{pmatrix}
\Psi^K=E\Psi^K
\end{align}
and
\begin{align}\label{DiracKprime}
v_F\begin{pmatrix}
0 &\hat{p}_x+i \hat{p}_y\\
\hat{p}_x-i \hat{p}_y &0
\end{pmatrix}
\Psi^{K'}=E\Psi^{K'}\,.
\end{align}
The fermi velocity is related to the lattice spacing 
and the overlap integral via $v_F=a_0\gamma_0 \sqrt{3}/2$.
The Dirac equations (\ref{DiracK}) and (\ref{DiracKprime}) have a particle-hole-symmetry, that is to say each positive-energy spinor $\Psi^{\mathfrak{p}}_E=[\psi_{A},\psi_{B}]^T$ has as a counterpart a negative-energy spinor 
$\Psi^{\mathfrak{h}}_E=\sigma_z \Psi^{\mathfrak{p}}_E$.
For an infinitely extended graphene sheet one 
finds the linear energy-momentum relation $E=\pm v_F\sqrt{p_x^2+p_y^2}$
whereas for a minimally coupled magnetic field
a splitting into Landau bands occurs \cite{G11}.


%

%
%

\section{Optical Conductivity}\label{seccond}

The conductivity of graphene without magnetic field has to leading 
order the frequency-independent value $\sigma=\pi \alpha_{QED}\approx 0.023$ \cite{PS08,KM16,MSW08}.
Applying a finite magnetic field to a graphene sheet leads to  an enhanced absorption for particular frequencies due to the 
peaked charge carrier densities around the Landau levels \cite{GS06,GSC06}.
One might argue that this enhancement alone could account for the large observed current.
However, the bulk modes will not contribute to the edge current which
motivates the investigation of the absorption into the dispersive edge modes in geometries with broken translational symmetry \cite{NFD96}.

The optical response of graphene which is subject to an 
external electric field $E_j(\mathbf{r},t)=E_j(\mathbf{r},\Omega)e^{-i\Omega t}$, i.e. a coherent laser field,
can be calculated perturbatively from the Kubo formula \cite{K57}.
The conductivity tensor $\sigma_{ij}$ links the external field 
to the induced current
\begin{align}\label{indcurrent}
J_i(\mathbf{r},\Omega)= \sum_j \int d^2r' \sigma_{ij}(\mathbf{r},\mathbf{r}',\Omega)E_j(\mathbf{r}',\Omega)
\end{align}
and can be expressed as current-current correlation function.
In order to evaluate the correlation function, it is convenient 
to employ thermal Green functions together with a subsequent 
analytical continuation, $\sigma_{ij}(\mathbf{r},\mathbf{r}',\Omega)=\sigma^T_{ij}(\mathbf{r},\mathbf{r}',i\Omega_l\rightarrow~ \Omega+i\delta)$.
Explicitly, the correlation function of the thermal Dirac currents 
$\hat{J}_i(\mathbf{r},\tau)=-qv_F \hat{\bar{\Psi}}(\mathbf{r},\tau)\gamma^i\hat{\Psi}(\mathbf{r},\tau)$
has the form
\begin{align}\label{thermsig}
&\sigma^T_{ij}(\mathbf{r},\mathbf{r}',i\Omega_l)=
-\frac{i}{\Omega}\int_0^\beta e^{i\Omega_l\tau}\langle T_\tau \hat{J}_i(\mathbf{r},\tau)\hat{J}_j(\mathbf{r}',0)\rangle\,.
\end{align}
The Dirac spinor $\hat{\Psi}$ contains the positive- and negative energy modes at both Dirac points.
From equations (\ref{DiracK}) and (\ref{DiracKprime}) we find the representation of the 
Dirac matrices, $\gamma^0~=~\mathrm{diag}(\sigma_z,\sigma_z)$, $\gamma^1=\mathrm{diag}(-i\sigma_y,i\sigma_y)$ and $\gamma^2=\mathrm{diag}(i\sigma_x,i\sigma_x)$.
Finally, the correlator (\ref{thermsig}) can expressed in terms of thermal Green functions,
\begin{multline}\label{Kubo}
\sigma^T_{ij}(\mathbf{r},\mathbf{r}',i\Omega_l)\hfill\\
=\frac{i(q v_F)^2}{\Omega}\int_0^\beta e^{i\Omega_l\tau}
\mathrm{Tr}\{\gamma^i\hat{G}(\mathbf{r},\mathbf{r}',\tau)\gamma^j\hat{G}
(\mathbf{r}',\mathbf{r},-\tau)\}
\end{multline}
which then allows the evaluation in terms of Matsubara sums \cite{mahan}.

We assume that the wavelength of the electric field is much larger than the cyclotron radius of an electron- or hole excitation and approximate $E_j(\mathbf{r},\Omega)\approx E_j(\Omega)$.
The diagonal components of the correlator determine then the conductivities for the polarizations of the electrical field
perpendicular and parallel to the graphene edge,
\begin{align}\label{cond}
\sigma_{\perp,\parallel}(\mathbf{r},\Omega)=\Re \lim_{\delta\rightarrow 0}\int d^2r'\sigma^T_{\perp,\parallel}(\mathbf{r},\mathbf{r}',\Omega+i\delta)\,. 
\end{align}
The evaluation of this expression requires the 
explict knowledge of Dirac spinor which we shall evaluate in the following for the graphene fold and the zigzag edge.

\subsection{Graphene fold}\label{conductivityfold}

We consider a graphene fold along the $y$-axis with a magnetic field pointing in $z$-direction.
The corresponding vector potential is minimally coupled to the Dirac equations 
(\ref{DiracK}) and (\ref{DiracKprime}) via $\hat{p}_y\rightarrow \hat{p}_y+q A(x)$.
The parallel momentum $k$ is preserved due to the 
translational invariance along the $y$-direction.
As consequence of the magnetic field, the particle and hole-excitations occupy Landau bands.
Alltogether, we can decompose the 
Dirac spinor into positive-energy ($\mathfrak{p}$) particle  modes and negative-energy ($\mathfrak{h}$) hole modes in the vicinity of both Dirac points according to
\begin{align}\label{spinorfold}
&\hat{\Psi}(\mathbf{r},t)=\frac{1}{2\pi\sqrt{2}}
\int d{k}\sum_{m>0}e^{iky}\\ 
&\Bigg[
\bigg\{\begin{pmatrix}
\Psi^{\mathfrak{p},K}_{k,m}(x)\\
0
\end{pmatrix}
\hat{a}^{K}_{m,k}+
\begin{pmatrix}
0\\
\Psi^{\mathfrak{p},K'}_{k,m}(x)
\end{pmatrix}
\hat{a}^{K'}_{m,k}\bigg\}e^{-i E_{m,k}t}\nonumber\\
&+\bigg\{\begin{pmatrix}
\Psi^{\mathfrak{h},K}_{k,m}(x)\\
0
\end{pmatrix}
\hat{b}^{\dagger K}_{m,k}+
\begin{pmatrix}
0\\
\Psi^{\mathfrak{h},K'}_{k,m}(x)
\end{pmatrix}
\hat{b}^{\dagger K'}_{m,k}\bigg\}e^{+i E_{m,k}t}
\Bigg]\nonumber\,,
\end{align}
where we used the fact that the energies of the eigenmodes 
coincide at both Dirac points.
%
%
From equations (\ref{DiracK}) and (\ref{DiracKprime}) we find 
that the spinor components of the Dirac points are related by
$\psi^{K'}_{A,k,m}=\psi^{K}_{B,k,m}$ and 
$\psi^{K'}_{B,k,m}=-\psi^{K}_{A,k,m}$.
Furthermore, the eigenvalue equations (\ref{DiracK}) 
and (\ref{DiracKprime}) can be decoupled which reduces the 
problem to the solution of the one-dimensional Schr\"odinger equation
%
$v_F^2[-\partial_x^2+V(x)]\psi^{K}_{A,k,m}=E_{m,k}^2 \psi^{K}_{A,k,m}$ 
%
with the effective potential $V(x)=(k+q\,A(x))^2+q\, \partial_x A(x)$.
%
%
%

Following the discussion in \cite{QS13}, a symmetric vector potential $A(-x)=A(x)$ 
gives rise to an additional symmetry which relates 
the spinor components via
$\psi^K_{B,k,m}(x)=-i \mathcal{P}_{k,m}\psi^K_{A,k,m}(-x)$
where  $\mathcal{P}_{k,m}=(-1)^{m+1}$ is called the pseudo-parity.
The lowest Landau band is labelled with $m=1$.
%
As a particular realization of a graphene fold with curvature radius $R$ in a constant magnetic field, we take the vector potential to be of the form
\begin{align}
A(x)=\begin{cases}
B_0 R \left(1-\cos \frac{x}{R}\right) &\mbox{if} \left|x\right|\leq \frac{\pi R}{2}  \\
B_0 R\left(  \left|\frac{x}{R}\right|-\frac{\pi}{2}+1\right)    &\mbox{if} \left|x\right|> \frac{\pi R}{2}
     \end{cases}\,.
\end{align}
%
%
For the sake of simplicity we shall take in the following the curvature radius to be equal to the magnetic length $\ell_{B_0}=1/\sqrt{qB_0}$.

\begin{figure}
\includegraphics[width=\columnwidth]{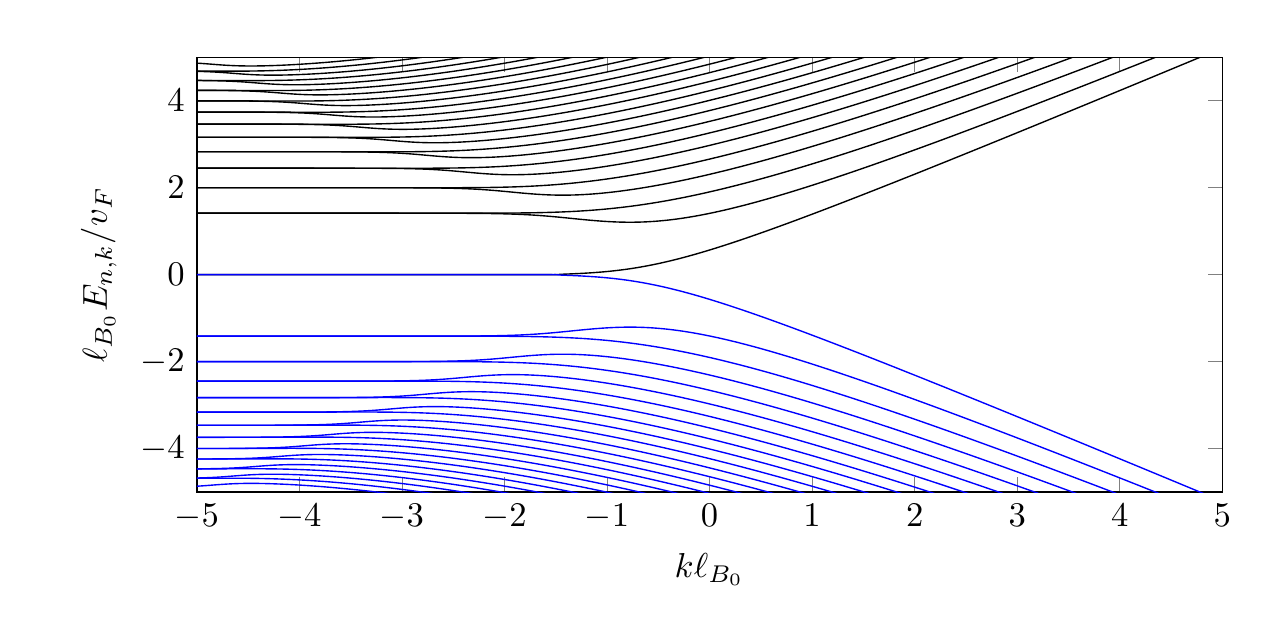} 
\caption{Particle-hole-symmetric energy spectrum for a graphene fold in a magnetic field. The bulk modes have a flat dispersion whereas the edge-modes we have $|dE/dk|\lesssim v_F $.}\label{energybandsfold}
\end{figure}

The eigensystem of the effective Schr\"odinger equation can be easily evaluated numerically, see Fig.~\ref{energybandsfold}. 
Nevertheless, we shall give analytical approximations 
for the eigenfunctions and energy bands which are relevant for our discussion in 
section \ref{mpfold}.
For large and positive $k$, the effective potential reads 
$V(x)\approx k^2+x^2k/(\ell_{B_0}^3)+x/\ell_{B_0}^3$,
hence the eigenfunctions are boundary modes and can be expressed in terms of the harmonic oscillator eigenfunctions 
%
\begin{align}\label{largeposkwave}
 \psi_{A,k,m}^K(x)\approx \phi_{m-1}((k\ell_{B_0})^{1/4}x/\ell_{B_0}),\quad m=1,2,...
\end{align}
The corresponding energy bands are then
\begin{align}\label{largeposk}
E_{m,k}\approx v_F\sqrt{2\sqrt{k \ell_{B_0}}(m-1/2)/\ell^2_{B_0}+k^2}\,.
\end{align}
Asymptotically the bands have a linear dispersion $\sim v_F k $ with an
offset depending on the particular Landau band.
For large but negative $k$, the effective potential takes the form
$V(x)\approx (k+|x|/\ell_{B_0}^2+(1-\pi/2)/\ell_{B_0})^2+\mathrm{sign}(x)/\ell_{B_0}^2$.
The corresponding eigenfunctions are dispersionless bulk modes which 
are well approximated by harmonic oscillator eigenfunctions.
For the states of positive pseudo-parity we have
\begin{align}
\psi_{A,k,1}^K(x)=&\phi_0\left(\frac{x}{\ell_{B_0}}-\ell_{B_0} k+\frac{\pi}{2}-1\right)\,\label{groundnegk}\\
\psi_{A,k,2n+1}^K(x)=&\frac{1}{\sqrt{2}}\bigg[\phi_{n+1}\left(\frac{x}{\ell_{B_0}}-\ell_{B_0} k+\frac{\pi}{2}-1\right)\nonumber\\
&-
\phi_n\left(-\frac{x}{\ell_{B_0}}-\ell_{B_0} k+\frac{\pi}{2}-1\right)\bigg]\,,
\end{align}
with the dispersionless Landau energies $E_{2n+1}=v_F\sqrt{2n}/\ell_{B_0}$ for $n=0,1,...$ .
Similarly we find for the states of negative pseudo-parity
\begin{align}
\psi_{A,k,2n}^K(x)=&\frac{1}{\sqrt{2}}\bigg[\phi_{n+1}\left(\frac{x}{\ell_{B_0}}-\ell_{B_0} k+\frac{\pi}{2}-1\right)\nonumber\\
&+
\phi_n\left(-\frac{x}{\ell_{B_0}}-\ell_{B_0} k+\frac{\pi}{2}-1\right)\bigg]\,,
\end{align}
and $E_{2n}=v_F\sqrt{2n}/\ell_{B_0}$ where $n=1,2,...$ .
To each Landau level with $n>0$ we have two states with 
opposite pseudo parity which are quasi-degenerate for sufficiently 
large negative $k$.
For a superposition $\tilde{\psi}^K_{A,k,m}$ of quasi-degenerate states one can employ the Dirac equation to construct 
a projection onto a state with definite pseudo-parity,
\begin{multline}
\frac{(-1)^{m+1}v_F}{E_{m,k}}\left[\partial_x +(k+q A(x))\right]
\tilde{\psi}^K_{A,k,m}(-x)+\tilde{\psi}^K_{A,k,m}(x)\\
\propto 
\psi^K_{A,k,m}(x)\,.
\end{multline}
For energies with $|k\ell_{B_0}|\lesssim 1$ the effective potential adopts the quartic form $V(x)\approx (k+x^2/(2\ell_{B_0}^3))^2+x/\ell_{B_0}^3$.
Although the eigenfunctions cannot be given in analytical form,
a reasonable approximation for the ground state wavefunction can be obtained from the variational ansatz
\begin{align}\label{groundsmallk}
\psi^K_{A,k,1}(x)=\left(\frac{a}{\pi}\right)^{1/4}\exp\left\{-\frac{a}{2 \ell_{B_0}}(x-x_0)^2\right\} 
\end{align}
with $a=1.09$ and $x_0=-0.58\ell_{B_0}
+0.47 k/\ell_{B_0}^2$.
The corresponding Landau band is dispersive can be approximated by $E_{1,k}\approx v_F\left(0.62/\ell_{B_0}+0.65 k\right)$,
cf.~Fig.~\ref{energybandsfold}.
As will be discussed in section \ref{mpfold}, these modes are of particular relevance for the charge current enhancement.

\begin{figure}[t]
\includegraphics[width=\columnwidth]{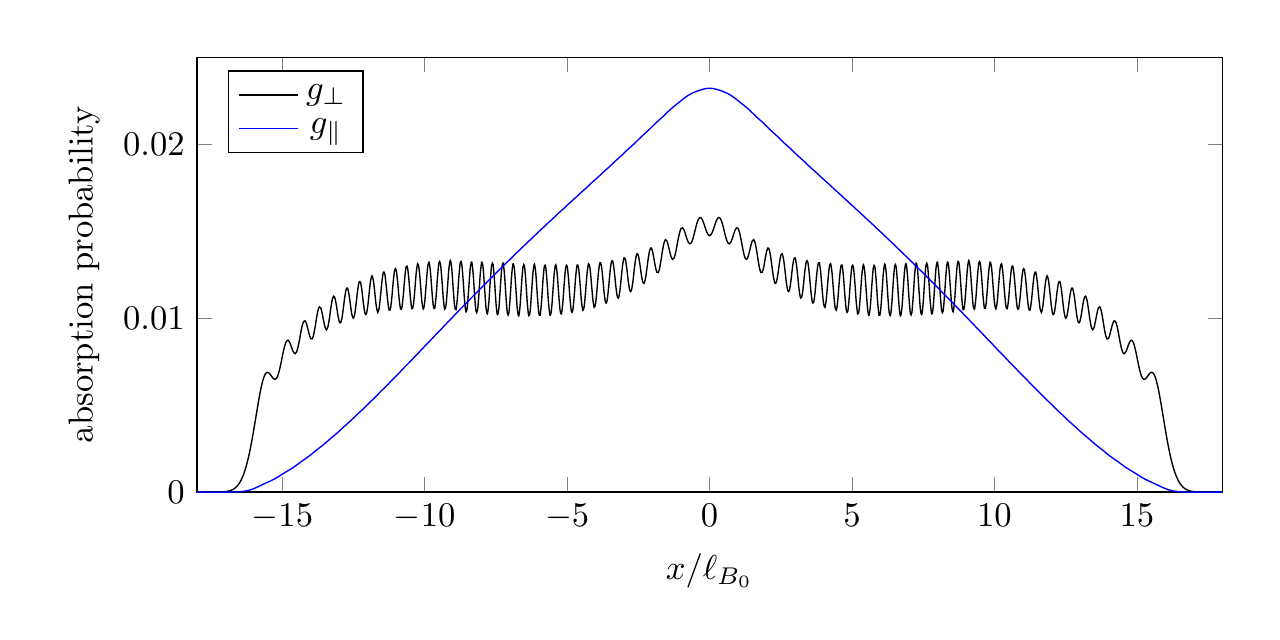} 
\caption{Absorption probability into the boundary modes for photons with 
polarization perpendicular and parallel to the fold. The values do not exceed the 
absorption of free graphene, $g_\mathrm{free}=\pi \alpha$.
The absorption probability vanishes when the distance between the absorbed photon and the 
edge exceeds the classical cyclotron radius of the charge carriers.
}\label{condfold}
\end{figure}

In terms of the eigenfunctions, one can evaluate the 
Green functions and subsequently the optical conductivities (\ref{cond}).
Employing the particle-hole-symmetry and the pseudo-parity,
we obtain at zero temperature for each spin degree of freedom 
%
\begin{align}\label{condxxyy}
&\sigma_{\perp,\parallel}(x,\Omega)=\frac{(qv_F)^2}{4 \Omega}\int dk
\sum_{m,l=1}^\infty  \delta(\Omega-E_{m,k}-E_{l,k})\nonumber\\
&\times\left[1+\lambda_{\perp,\parallel}(-1)^{l+m}\right]\int dx'
\psi^K_{A,k,m}(-x')\psi^K_{A,k,l}(x')
\nonumber\\
&\times \psi^K_{A,k,m}(-x)\psi^K_{A,k,l}(x)+(K\rightarrow K')\,.
\end{align}
Here we have $\lambda_\perp=1$ ($\lambda_\parallel=-1$) for the conductance perpendicular (parallel) to the fold.
We infer from (\ref{condxxyy}) that the conductivity perpendicular to the fold involves only transitions between states of equal pseudo-parity and the states with opposite pseudo-parity define the  conductivity parallel to the edge.

The absorption probability $g$ is the ratio of absorbed energy flux, $W_\mathrm{abs}=\sum_i J_iE_i$ and incident energy flux $W_\mathrm{inc}=\sum _i E_iE_i/(4\pi)$.
Summing over the spin degrees of freedom gives then the relation 
between conductivity and absorption probability, $g_{\perp,\parallel}(x,\Omega)=8\pi \sigma_{\perp,\parallel}(x,\Omega)$.

\begin{figure}
\includegraphics{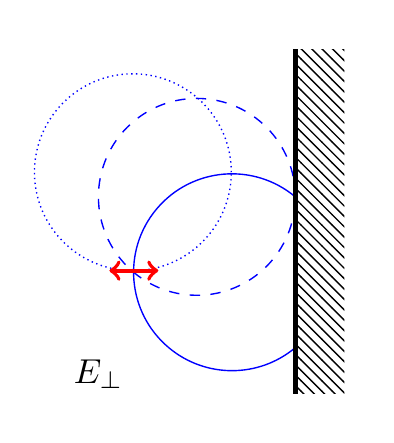}
\includegraphics{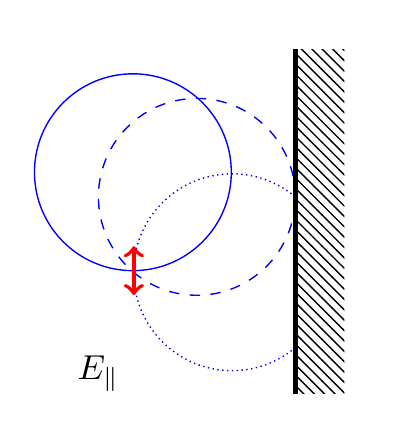}
\caption{
Left side: Perpendicular polarized photons 
generate particle-hole pairs which are propagating on circular trajectories.
Most likely is the propagation on the solid 
circle whereas the trajectories of the dashed and dotted circles are  less probable.
We conclude that for this polarization most of the generated particle-hole pairs will be reflected at the graphene edge for distances $\lesssim 2r_\mathrm{cyc}$, cf. $g_\perp$ in Fig.~\ref{condfold}.
Right side: The same argument shows that for parallel polarized photons which are absorbed at a distance $\sim 2r_\mathrm{cyc}$ from the edge, the reflection of the
generated charge-carriers at the edge is unlikely.
A significant absorption can only be expected when the photon is absorbed at a distance $\sim r_\mathrm{cyc}$ to the edge which explains the gradual increase of $g_\parallel$ in Fig.~\ref{condfold}.
}\label{absorptionscetch}
\end{figure}

The polarization- and distance dependent absorption 
probability into the dispersive edge channels is depicted in Fig.~\ref{condfold}.
We choose the photon energy to be $\Omega=17v_F/\ell_{B_0}$
at a magnetic field strength of $B=5T$ which 
corresponds to a photon energy of $\Omega=1$eV.
In a semiclassical picture, a generated electron-hole pair would then propagate on circular trajectories with radius $r_\mathrm{cyc}=(\Omega/2) \ell^2_{B_0}/v_F=8.5\ell_{B_0}$
Keeping this in mind, we can interpret the qualitative behavior of 
the $g_\perp$ and $g_\parallel$.
For a photon which is polarized perpendicular to the fold, the transition 
matrix elements are proportional to the momentum component $k_\parallel$, i.e. $\bar{\Psi}_{E}\gamma^1\Psi_{-E}\propto 
k_\parallel/\sqrt{k_\parallel^2+k_\perp^2}$.
Thus, as shown in the left panel of Fig.~\ref{absorptionscetch}, the charge carriers tend to propagate parallel to the fold before the magnetic field forces them on circular trajectories.
As consequence, many of the trajectories will intersect with the graphene fold if the particle-hole
pair is generated at a distance $x\lesssim 2r_\mathrm{cyc}$.
This explains the 
rather sudden increase of $g_\perp(x)$ in Fig.~\ref{condfold}.
In contrast, an absorbed photon being polarized parallel to the fold is likely to generate particle-hole pairs propagating perpendicular to the fold before their trajectories 
are bend by the magnetic field, $\bar{\Psi}_{E}\gamma^2 \Psi_{-E}\propto 
k_\perp/\sqrt{k_\parallel^2+k_\perp^2}$.
From the right panel in Fig.~\ref{absorptionscetch} we infer that it is rather unlikely that charge carriers are reflected from the fold if they are generated at a distance $x\lesssim 2r_\mathrm{cyc}$.
This in turn explains the gradual increase of~$g_\parallel(x)$ towards the fold in Fig.~\ref{condfold}.

%

\subsection{Zigzag boundary}

\begin{figure}
\includegraphics[width=\columnwidth]{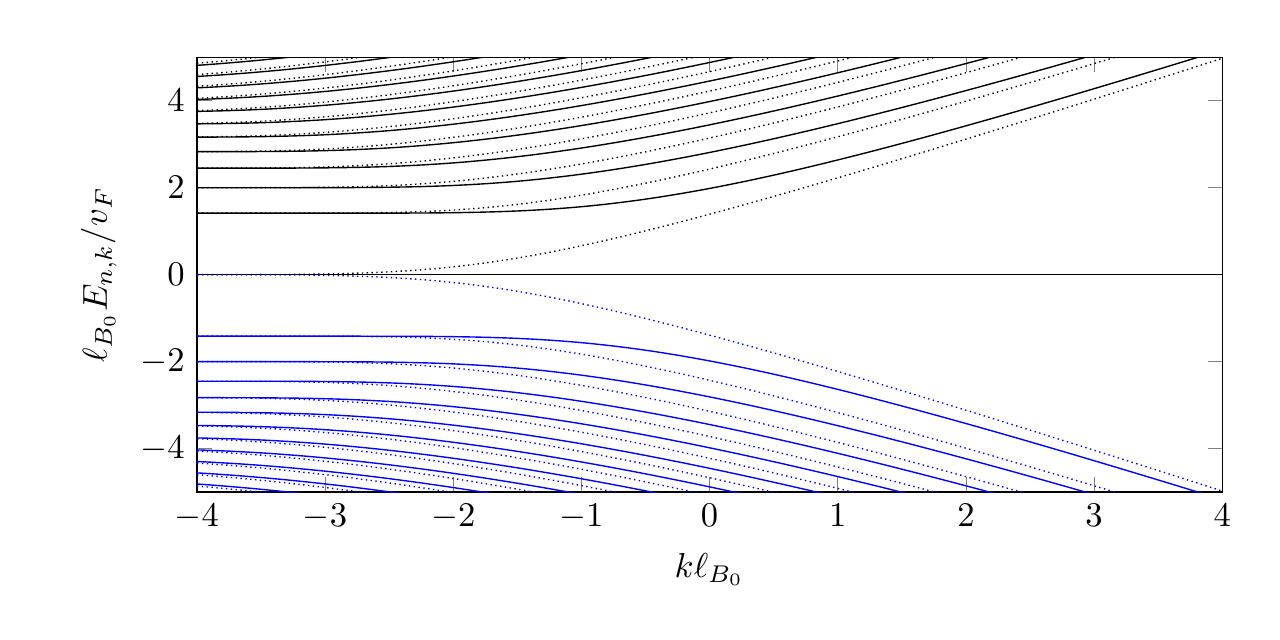}
\caption{Energybands for a graphene sheet with zigzag edge in a magnetic field.
The dotted curves correspond to the energybands for the modes in the vicinity 
of the $K'$-point whereas the energybands for the second Dirac point $K$ 
are represented by solid lines. 
The boundary condition induces a zero energy mode at $K$.}\label{energyzigzag}
\end{figure}

As noted before, the coordinate system for the tight-binding equations (\ref{tight1}) and (\ref{tight2}) was selected such that 
a zigzag boundary is parallel to the $x$-axis.
Therefore we take the non-vanishing component of the 
vector potential to be in $x$-direction, $A_x(y)=-B_0y$.
As in the fold-geometry, the two Dirac points can be treated separately.
After decoupling the Dirac equations (\ref{DiracK}) and (\ref{DiracKprime}) one obtains the effective Schr\"odinger equations
\begin{align}\label{effschrzigzag1}
[-\partial_y^2+(k-q\,B_0y)^2+qB_0]\psi^{K}_{A,k,m}=\frac{E_{m,k}^2}{v_F^2} \psi^{K}_{A,k,m} 
\end{align}
and
\begin{align}\label{effschrzigzag2}
[-\partial_y^2+(k-q\,B_0y)^2-qB_0]\psi^{K'}_{A,k,m}=\frac{{E^{'2}_{m,k}}}{v_F^2} \psi^{K'}_{A,k,m}\,.
\end{align}
As the graphene sheet possesses a zigzag boundary, 
we can choose  $A$-sublattice.
Therefore the boundary condition translates to $\psi^{K}_{A,k,m}(y=0)=\psi^{K'}_{A,k,m}(y=0)$=0.
%
The energy bands are shown in Fig.~\ref{energyzigzag} and the 
corresponding eigenfunctions can be expressed in terms of parabolic cylinder functions \cite{PSim08,ALL07}.
Around the $K$-point there exists a zero-energy mode with one vanishing spinor component,
$\psi^{K}_{A,k,0}(y)=0$ and $\psi^{K}_{B,k,0}(y)\sim \exp[-(qB_0 y-k)^2/(2qB_0)]$.
Decomposing the Dirac spinor into eigenmodes which are labelled by the parallel momentum, the Landau band and the Dirac point, we obtain
%

 \begin{align}\label{spinorzigzag}
& \hat{\Psi}(\mathbf{r},t)=\frac{1}{2\pi\sqrt{2}}
 \int d{k}\sum_{m=0}e^{ikx}\nonumber\\ 
& \Bigg[
\begin{pmatrix}
0\\
\Psi^{\mathfrak{p},K'}_{k,m}(y)
\end{pmatrix}
\hat{a}^{K'}_{m,k}e^{-i E^{K'}_{m,k}t}
+
\begin{pmatrix}
0\\
\Psi^{\mathfrak{h},K'}_{k,m}(y)
\end{pmatrix}
\hat{b}^{\dagger K'}_{m,k}e^{+i E^{K'}_{m,k}t}
\Bigg]\nonumber\\
%
&+\frac{1}{2\pi\sqrt{2}}
 \int d{k}\sum_{m>0}e^{ikx}\nonumber\\ 
& \Bigg[
\begin{pmatrix}
\Psi^{\mathfrak{p},K}_{k,m}(y)\\
0
\end{pmatrix}
\hat{a}^{K}_{m,k}e^{-i E^{K}_{m,k}t}
+
\begin{pmatrix}
\Psi^{\mathfrak{h},K}_{k,m}(y)\\
0
\end{pmatrix}
\hat{b}^{\dagger K}_{m,k}e^{+i E^{K}_{m,k}t}
\Bigg]\nonumber\\
&+\frac{1}{2\pi}
 \int d{k}\,e^{ikx} 
 \begin{pmatrix}
\Psi^{K}_{k,0}(y)\\
0
\end{pmatrix}
\hat{a}^{K}_{0,k}
\end{align}
The components of $\Psi^{\mathfrak{h},K,K'}_{k,m}(y)$
and $\Psi^{\mathfrak{p},K,K'}_{k,m}(y)$
can taken to be real (cf.~Eqs.~(\ref{DiracK}).
After some algebra we obtain
for the conductance
%
%
\begin{align}\label{condzigzagperppar}
&\sigma_{\perp,\parallel}(y,\Omega)= \frac{(qv_F)^2}{4\Omega}\int dk \int dy'\sum_{l=1}^\infty\nonumber\\
&\times\bigg[\sum_{m=1}^\infty\delta(\Omega-E^K_{m,k}-E^K_{l,k})\psi^K_{A,k,l}(y')\psi^K_{B,k,m}(y')\nonumber\\
&\times [\psi^K_{A,k,l}(y)\psi^K_{B,k,m}(y)+\lambda_{\perp,\parallel}\psi^K_{A,k,m}(y)\psi^K_{B,k,l}(y)]\nonumber\\
&+\delta(\Omega-E^K_{l,k})
\psi^K_{A,k,l}(y')\psi^K_{B,k,0}(y')\psi^K_{A,k,l}(y)\psi^K_{B,k,0}(y)\bigg]\nonumber\\
+&
\frac{(qv_F)^2}{4\Omega}\int dk \int dy'\sum_{l,m=0}^\infty\nonumber\\
&\times\bigg[\delta(\Omega-E^{K'}_{m,k}-E^{K'}_{l,k})\psi^{K'}_{A,k,l}(y')\psi^{K'}_{B,k,m}(y')\nonumber\\
&\times [\psi^{K'}_{A,k,l}(y)\psi^{K'}_{B,k,m}(y)+\lambda_{\perp,\parallel}\psi^{K'}_{A,k,m}(y)\psi^{K'}_{B,k,l}(y)]
\end{align}
As for the fold geometry, cf.~(\ref{condxxyy}), the conductivity is determined by the generation of particle-hole 
pairs at $K$ and $K'$.
However, for the zigzag boundary, the excitations around the Dirac points contribute differently to 
the absorption probability.
In particular the presence of the zero energy mode
at the Dirac point $K$ permits the absorption of 
the photon energy into a single particle- or hole excitation,
see equation (\ref{condzigzagperppar}).
This contribution is independent of the polarization. 
%
The qualitative behavior of the absorption probabilities $g_\perp$ 
and $g_\parallel$ can be justified with the same arguments as 
before.

\begin{figure}[h]
\includegraphics[width=\columnwidth]{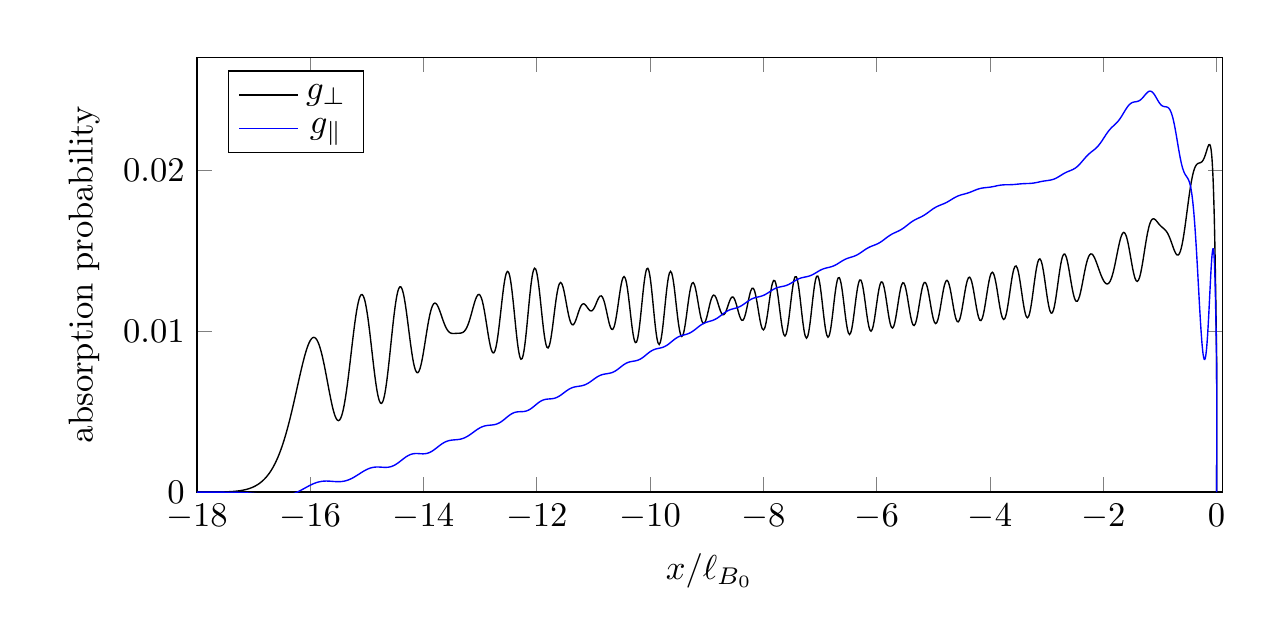} 
\caption{Absorption probability into the dispersive edge modes for 
the polarizations perpendicular and parallel to the zigzag edge.
The curves show the same qualitative behavior as the
absorption probabilities around a graphene fold.
The peaks in the vicinity of $x=0$ are mainly due to the presence of the 
zero energy mode.
}\label{condzigzag}
\end{figure}

From Figs.~(\ref{condfold}) and (\ref{condzigzag}) we infer that the maximum absorption probability 
into the edge modes does not exceed the absorption of monolayer graphene, $g_0=\pi\alpha_\mathrm{QED}$.
Note that we selected $\Omega$ to be off-resonant such that the absorption into the bulk modes did not contribute.
Alltogether we conclude that the large current observed in \cite{SKG17}
cannot be traced back to an enhanced absorption.
As possible explanation for this effect we shall consider 
estimates for charge carrier multiplication rates due to 
impact ionization.

\section{Secondary particle-hole pair creation}\label{secparthole}

\begin{figure}
\includegraphics{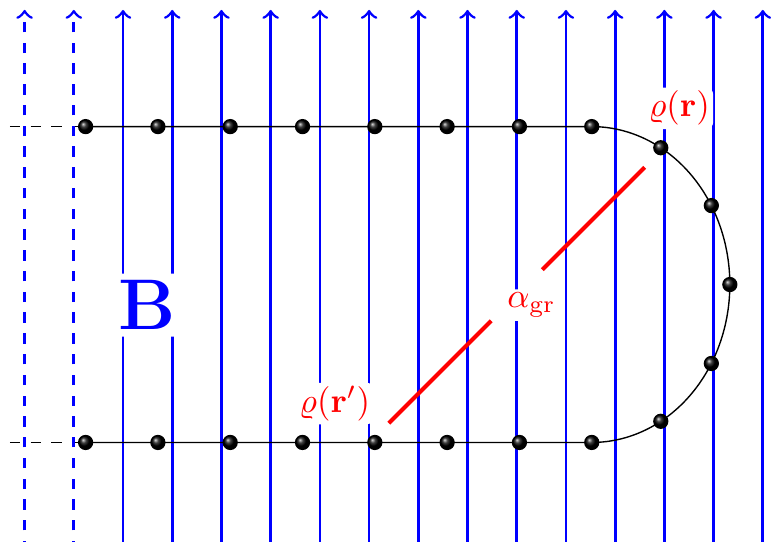}
\caption{Sketch of the graphene fold in an 
external magnetic field. Secondary 
particle-hole pairs are generated via the Coulomb 
interaction.
}\label{foldfigure}
\end{figure}

Auger processes have been studied thoroughly
for translational invariant graphene sheets \cite{TBC13,WKM10,WKM14,MWB}.
The energy-momentum conservation together with the linear 
dispersion relation restricts the phase space for interactions to one dimension, i.e. only collinear 
processes are allowed \cite{TBC13}.
In contrast, a graphene edge breaks the 
translational invariance and opens up a finite 
phase-space volume for carrier multiplication.

Our analysis will reside on the bare Coulomb Hamiltonian
\begin{align}
\hat{H}_\mathrm{Coulomb}=\frac{q^2}{2}\int d^2r \int d^2{r'}\frac{\hat{\varrho}(\mathbf{r})\hat{\varrho}(\mathbf{r}')}{4 \pi\epsilon_0|\mathbf{r}-\mathbf{r}'|}\,,
\end{align}
where the electron densities can be  written in terms of the Dirac fields 
as
\begin{align}
\hat{\varrho}(\mathbf{r})=\sum_\sigma\left(\big[\hat{\Psi}_\sigma^{K}(\mathbf{r})\big]^\dagger \hat{\Psi}_\sigma^{K}(\mathbf{r})+
\big[\hat{\Psi}_\sigma^{K'}(\mathbf{r})\big]^\dagger \hat{\Psi}_\sigma^{K'}(\mathbf{r})\right)\,.
\end{align}
Here we employed the expressions (\ref{Alattice}) and (\ref{Blattice})
and neglected rapidly oscillating terms $\sim e^{i(\mathbf{K}-\mathbf{K}')\cdot \mathbf{r}} $.
The label $\sigma$ denotes the spin degree of freedom 
which was suppressed in the previous sections.

We emphasize that our study is solely based on Fermi's Golden rule.
Our analysis for carrier multiplication around the graphene edge shall give only a rough order-of-magnitude estimate 
for the charge carrier multiplication whereas the actual scattering rates 
are likely to be smaller.
In fact, the dielectric constant will be diminished due to the substrate material or electronic screening \cite{A06,HS07}.
However, as consequence of the vanishing density of states, electronic screening at the Dirac point is rather 
inefficient \cite{A06}. 
%
%
%

We consider Auger-type inelastic 
scattering of an incoming electron, $|\mathrm{in}\rangle=|k_\mathrm{in},n_\mathrm{in}\rangle$, 
to an outgoing electron, $|\mathrm{out}\rangle=|k_\mathrm{out},n_\mathrm{out}\rangle$, while creating 
an electron-hole pair $|\mathrm{part},\mathrm{hole}\rangle=|k_\mathrm{part},n_\mathrm{part},k_\mathrm{hole},n_\mathrm{hole}\rangle $.
%
Using standard first order perturbation theory, we find
for the transition matrix elements 
\begin{multline}\label{matelement}
M(\mathrm{in}\rightarrow \mathrm{out}, \mathrm{part}, \mathrm{hole} )\\
=-i\int dt \langle \mathrm{out},\mathrm{part},\mathrm{hole}|\hat{H}_\mathrm{Coulomb}|\mathrm{in}\rangle\,.
\end{multline}
Summing the absolute square of (\ref{matelement})
over all final states gives the leading order result
for the impact ionization rate.
If we assume, that the momentum of the incoming particle
is in the vicinity of the Dirac point $K$ and 
the secondary particle-hole pair is generated around 
$K'$, the probability per unit time reads
\begin{multline}\label{kkprime}
\frac{\mathcal{P}^{K,\sigma\rightarrow K',\lambda}}{T}=\frac{v_F\, \alpha_\mathrm{graphene}^2}{8\pi }\int dk_\mathrm{out}
\mathcal{H}^{K,\sigma\rightarrow K',\lambda}(k_\mathrm{out})
\end{multline}
where the integrand 
\begin{align}
\mathcal{H}^{K,\sigma\rightarrow K',\lambda}(k_\mathrm{out})=
\frac{v_F}{|\frac{d\mathcal{F}}{d k_\mathrm{part}}|}\left|\mathcal{I}^{K,\sigma\rightarrow K',\lambda}\right|^2 \bigg|_{\mathcal{F}=0}
\end{align}
is determined by an overlap integral specifying 
the Coulomb interaction between charge densities,
\begin{align}\label{overlapkkprime}
\mathcal{I}^{K,\sigma\rightarrow K',\lambda}
=\int dx \int dx' \left[\Psi^{\mathfrak{p},K'}_{k_\mathrm{part},n_\mathrm{part}}(x)\right]^\dagger
\Psi^{\mathfrak{h},K'}_{k_\mathrm{hole},n_\mathrm{hole}}(x)\nonumber \\
\times K_0\left(|k_\mathrm{in}-k_\mathrm{out}|D(x,x')\right)
\left[\Psi^{\mathfrak{p},K}_{k_\mathrm{out},n_\mathrm{out}}(x')\right]^\dagger
\Psi^{\mathfrak{p},K}_{k_\mathrm{in},n_\mathrm{in}}(x')
\end{align}
%
%
and a weight factor $d\mathcal{F}/dk_\mathrm{part}$, see below.
The modified Bessel function of second kind $K_0$ in (\ref{overlapkkprime}) arises from integrating out the 
direction parallel to the edge.
Its argument contains the momentum difference between incoming and outgoing particle and the function $D(x,x')$ that measures the 
distance of the charge densities perpendicular to the fold.
Two variables can be eliminated using the momentum conservation, $k_\mathrm{in}-k_\mathrm{out}=k_\mathrm{part}-k_\mathrm{hole}$, and the energy conservation, $\mathcal{F}=E_{n_\mathrm{in},k_\mathrm{in}}-E_{n_\mathrm{out},k_\mathrm{out}}-E_{n_\mathrm{part},k_\mathrm{part}}-E_{n_\mathrm{hole},k_\mathrm{hole}}\equiv0$.
The weight factor $d\mathcal{F}/dk_\mathrm{part}$ in (\ref{kkprime}) ensures the local reparametrization invariance of the 
integral.
The reparametrization invariance for the whole integration
domain does not exists in general since
$dk_\mathrm{out}/dk_\mathrm{part}$ can become 
singular due to energy-momentum conservation.
Nevertheless, it is always possible to find a parametrization for $\mathcal{H}$ which governs the complete integration domain. 

%
The matrix-elements for impact ionization in the same Dirac valley
but opposite spin are analogous to (\ref{kkprime}) with $K'\rightarrow K$ or $K\rightarrow K'$.
In contrast, when the spin and the valley index are the same,
the outgoing electron is indistinguishable from the generated particle.
Therefore the probability $\mathcal{P}^{K,\sigma\rightarrow K,\sigma}$ is now determined by overlap integrals which 
satisfy an exchange symmetry,
\begin{align}\label{kktransition}
&\mathcal{I}^{K,\sigma\rightarrow K,\sigma}
=\int dx \int dx' \left[\Psi^{\mathfrak{p},K}_{k_\mathrm{part},n_\mathrm{part}}(x)\right]^\dagger
\Psi^{\mathfrak{h},K}_{k_\mathrm{hole},n_\mathrm{hole}}(x)\nonumber \\
&\times K_0\left(|k_\mathrm{in}-k_\mathrm{out}|D(x,x')\right)
\left[\Psi^{\mathfrak{p},K}_{k_\mathrm{out},n_\mathrm{out}}(x')\right]^\dagger
\Psi^{\mathfrak{p},K}_{k_\mathrm{in},n_\mathrm{in}}(x')\nonumber\\
&-(k_\mathrm{out},n_\mathrm{out}\leftrightarrow k_\mathrm{part},n_\mathrm{part})\,.
\end{align}
Before we discuss the specific settings of a 
graphene fold and a zigzag boundary, two remarks 
are in order.
First, the leading-order perturbation theory may not be very accurate since the 
dimensionless expansion parameter is larger 
than unity, $\alpha_\mathrm{graphene}=\alpha_\mathrm{QED}{c}/{v_F}\approx2.2$.
%
%
Second, we will not consider Auger recombination and assume that the generated charge carriers reach the bond contacts of the graphene boundary before particle-hole annihilation occurs.
Particle-hole annihilation can in principal be considered within a Boltzmann-equation approach \cite{TBC13}.
However, we expect the recombination rates to be negligible for sufficiently small electron-hole densities, see also \cite{R07}.

%


\subsection{Graphene fold}\label{mpfold}
A graphene fold breaks translational invariance without terminating the graphene sheet.
As in section \ref{conductivityfold}, we assume that the magnetic length $\ell_\mathrm{B} $ and the fold radius $R$ to be equal.
The spatial separation $D(x,x')$ of the charge carrier densities can be deduced 
by simple geometric considerations from Fig.~\ref{foldfigure}.

An exact selection rule occurs due to the pseudo parity which was briefly discussed in section \ref{conductivityfold}.
As consequence, the integrals 
$\mathcal{I}^{K,\sigma\rightarrow K,\lambda}$ and $\mathcal{I}^{K,\sigma\rightarrow K',\lambda}$ vanish identically 
unless the sum of the Landau level indices, $n_\mathrm{in}+n_\mathrm{out}+n_\mathrm{part}+n_\mathrm{hole}$, equals an odd integer.
From the allowed transitions, only a few decay channels dominate the process, whereas most of the channels will be suppressed 
by several orders of magnitude.
%
%
If the number of nodes of the wavefunctions $\Psi^{\mathfrak{p},K}_{k_\mathrm{in},n_\mathrm{in}}$
and $\Psi^{\mathfrak{p},K}_{k_\mathrm{out},n_\mathrm{out}}$
are very different from each other, the wave functions are nearly orthogonal which renders the overlap integral (\ref{overlapkkprime}) exponentially small.
The same is true for $\Psi^{\mathfrak{p},K'}_{k_\mathrm{part},n_\mathrm{part}}$
and $\Psi^{\mathfrak{p},K'}_{k_\mathrm{hole},n_\mathrm{hole}}$.
Therefore, the decay channels between the Dirac points $K$ and $K'$ can usually be neglected unless $n_\mathrm{part}\sim n_\mathrm{hole}$
and $n_\mathrm{in}\sim n_\mathrm{out}$.

Furthermore, a strong oscillation of the integrands will render the  overlap integrals exponentially 
small.
In order to keep the total number of nodes of the integrand in equation (\ref{overlapkkprime})
as small as possible, we can conclude that for an 
 incoming particle at Landau level $n_\mathrm{in}$
the following approximate 
selection rule applies:
\begin{align}\label{selrule}
n_\mathrm{out}\in\{n_\mathrm{in}, n_\mathrm{in}\pm 1\} \text{ and } n_\mathrm{part}\sim n_\mathrm{hole}\sim 1\,. 
\end{align}
For transitions within the same Dirac point, we obtain also relevant contributions for $n_\mathrm{out}\sim n_\mathrm{hole}\gg n_\mathrm{in}\sim n_\mathrm{part}$, a direct consequence 
of the exchange symmetry, see equation (\ref{kktransition}).

We illustrate our findings and consider the decay process  
of an incoming particle 
with $n_\mathrm{in}=5$ and $k_\mathrm{in}=10/\ell_{B_0}$.
Assuming a magnetic field $B_0=5T$, this would correspond to an 
initial energy of $E_{n_\mathrm{in},k_\mathrm{in}}=0.7$ eV.
For transitions involving states of both Dirac points, we listed 
the rates for various decay channels in Table \ref{tabKKprime}.
%
Although the applicability of first order perturbation 
theory should be doubted and not every generated charge carrier 
will contribute to the overall current, see below, we see from Table \ref{tabKKprime} that each of the largest decay channels generates between 80 and 90 particle-hole pairs within one picosecond which is about 10 particle-hole pairs within the 
distance of a classical cyclotron radius.
%

\begin{figure}[h]
\includegraphics[width=\columnwidth]{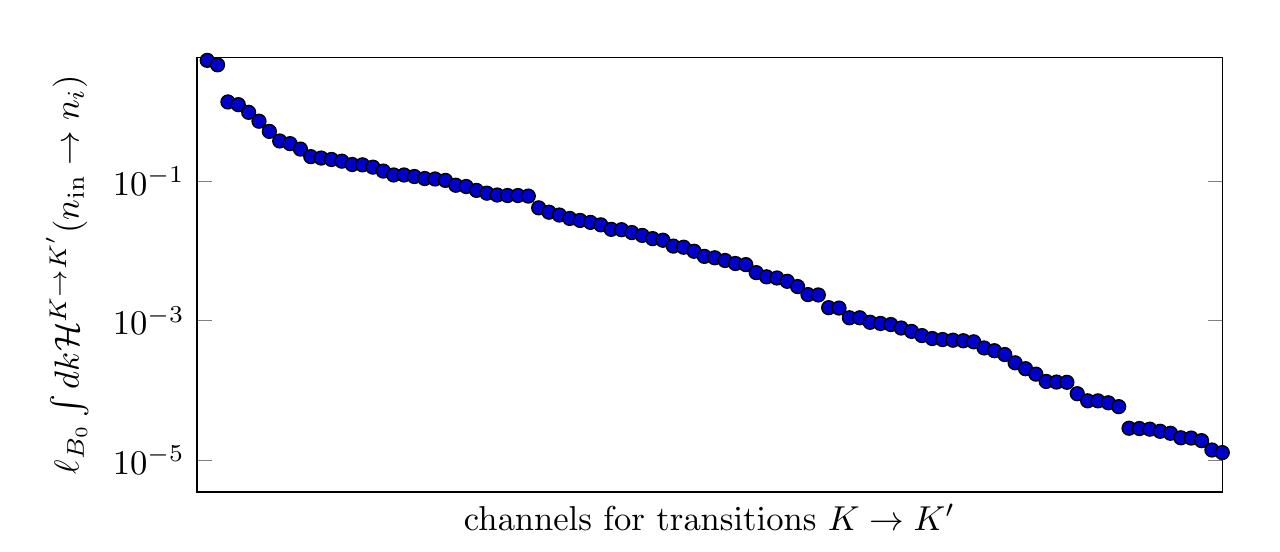} 
\caption{Overlap integrals which determine the decay probabilities 
per unit time.
Here we only consider the channels for which the incoming particle 
is in the vicinity of the $K$-point and the outgoing particle-hole pair in generated around the $K'$-point.
The corresponding rates for the two largest integrals are listed 
in Table \ref{tabKKprime}.
As can be seen here, the remaining channels are strongly suppressed.}\label{channelsKKprime}
\end{figure}

%
%
Fig.~\ref{channelsKKprime} shows the magnitude of the integrals which determine 
the decay channels $\mathcal{P}^{K,\sigma\rightarrow K',\lambda}(n_\mathrm{in}\rightarrow
n_\mathrm{out},n_\mathrm{part},n_\mathrm{hole})$.
Transitions which do 
not fulfill the relation (\ref{selrule}) are strongly suppressed in 
comparison with the dominating decay channels.
\begin{table}[h!]
\begin{center}
\begin{tabular}{l|c|c|c} 
$n_\mathrm{out}$ & $n_\mathrm{part}$ & $n_\mathrm{hole}$& $\mathcal{P}/T$ in s$^{-1}$ \\
\hline
\hline
4 & 1 & 1 & $ 9.4\cdot 10^{13}$\\
6 & 1 & 1 & $ 8.1\cdot 10^{13}$\\
1 & 4 & 1 & $ 1.5\cdot 10^{12}$\\
1 & 6 & 1 & $ 4.5\cdot 10^{11}$
\end{tabular}
\caption{The first two rows specify the largest decay rates for transitions $K\rightarrow K'$ for an incoming particle with $n_\mathrm{in}=5$ and $k_\mathrm{in}~=~10/\ell_{B_0}$. The decay rates in the third and fourth line 
are at least one magnitude smaller since they do not fulfill (\ref{selrule}).}\label{tabKKprime}
\end{center}
\end{table}

\begin{figure}[h]
\includegraphics[width=\columnwidth]{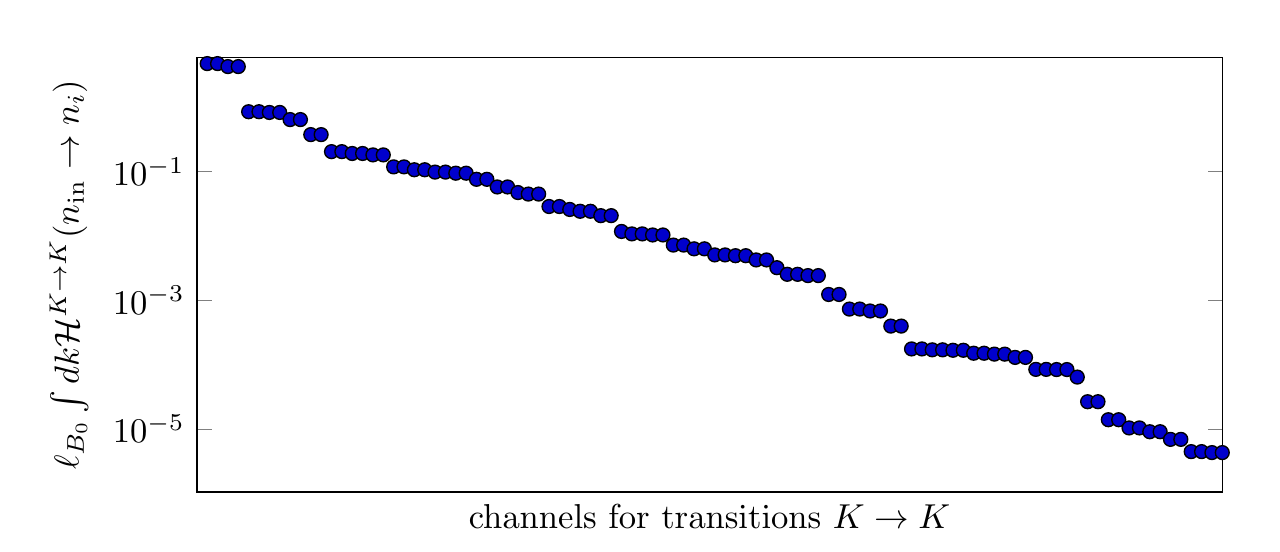} 
\caption{Overlap integrals determining the decay probabilities per unit time in the vicinity of one Dirac point time sorted by size on a logarithmic scale. As consequence of the exchange symmetry, all matrix elements occur in pairs.}\label{channelsKK}
\end{figure}

For transitions in the vicinity of one Dirac point, 
we listed the dominating integrals in 
Table \ref{tabKK}.
Here, all rates come in pairs which is a direct consequence 
of the exchange symmetry of (\ref{kktransition}),
from which follows that $\mathcal{P}^{K,\sigma\rightarrow K,\sigma}(n_\mathrm{out},n_\mathrm{part})=\mathcal{P}^{K,\sigma\rightarrow K,\sigma}(n_\mathrm{part},n_\mathrm{out})$.
Comparing the corresponding rates of Table \ref{tabKKprime}
and \ref{tabKK} we find $\mathcal{P}^{K,\sigma\rightarrow K,\sigma}\lesssim \mathcal{P}^{K,\sigma\rightarrow K',\sigma}$.
This small suppression in comparison to the $K\rightarrow K'$-rates 
originates from destructive interference which are 
rather small unless $n_\mathrm{out}\sim n_\mathrm{part}$,
cf.~equations (\ref{selrule}) and (\ref{kktransition}).
In Fig.~\ref{channelsKK}, we show the matrix elements 
determing the $K\rightarrow K$-transitions sorted by size.
\begin{table}
\begin{center}
\begin{tabular}{l|c|c|c} 
$n_\mathrm{out}$ & $n_\mathrm{part}$ & $n_\mathrm{hole}$& $\mathcal{P}/T$ in $ s^{-1}$\\
\hline
\hline
4 & 1 & 1 & $ 8.3\cdot 10^{13}$\\
1 & 4 & 1 & $ 8.3\cdot 10^{13}$\\
6 & 1 & 1 & $ 7.5\cdot 10^{13}$\\
1 & 6 & 1 & $ 7.5\cdot 10^{13}$
\end{tabular}
\caption{Decay rates of the dominating channels for an incoming particle with $n_\mathrm{in}=5$ and $k_\mathrm{in}=10/\ell_{B_0}$. The rates are symmetric w.r.t the interchange of $n_\mathrm{out}$ and $n_\mathrm{part}$.}\label{tabKK}
\end{center}
\end{table}

Summing over the spin configurations, both Dirac points and all final configurations for the outgoing particle and the generated particle-hole 
pair, we find for our example the total impact ionization rate
\begin{align}\label{examplerate}
\frac{\mathcal{P}_\mathrm{total}}{T}&=\frac{1}{T}
\sum_\lambda\sum_{n_\mathrm{I}}\left(
3\mathcal{P}_\mathrm{n_I}^{K,\sigma\rightarrow K',\lambda}+\mathcal{P}_\mathrm{n_I}^{K,\sigma\rightarrow K,\lambda}\right)\nonumber\\
&\approx1.4\cdot 10^{15}\mathrm{s}^{-1}
\end{align}
which corresponds to about 180 generated particle-hole 
pairs within a distance of one cyclotron radius.
Although some of the secondary particles and holes are generated in the bulk and do not contribute to the total current, see below, our estimate shows the efficiency 
of the charge carrier multiplication.

\begin{figure}[h]
\includegraphics[width=\columnwidth]{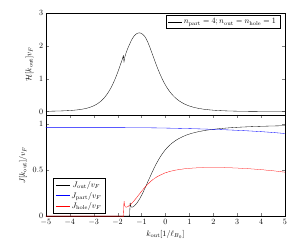} 
\caption{Upper panel: Value of $\mathcal{H}^{K,\sigma\rightarrow K,\sigma}$ as function 
of $k_\mathrm{out}$.
Lower panel: Currents related to the outgoing particle and the generated 
particle-hole pair. The cusps in the plots around $k_\mathrm{out}=-1.5/l_{B_0}$ originate from the slope of the lowest energy band in our model, cf. Fig.~\ref{energybandsfold}.}\label{slopesandprob5141}
\end{figure}

In our example, one of the dominating channels in the pair production 
process with transitions $K\rightarrow K$ is specified by the 
quantum numbers $n_\mathrm{in}=5$, $n_\mathrm{out}=1$, $n_\mathrm{part}=4$ and $n_\mathrm{hole}=1$.
The corresponding integrand $\mathcal{H}^{K,\sigma\rightarrow K,\sigma}$ for this decay channel is shown in the upper panel of 
Fig.~\ref{slopesandprob5141}.
In the lower panel in Fig.~\ref{slopesandprob5141} we see the currents $J=dE/dk$ for the outgoing particle and the generated particle hole-pair.
For $k_\mathrm{out}\lesssim-1.5/l_{B_0}$, the whole energy of the incoming 
particle is transferred to the generated particle whereas the generated hole 
and the outgoing particle are bulk modes with zero  energy.
In contrast, for $k_\mathrm{out}\gtrsim-1.5/l_{B_0}$, the outgoing particle and the generated hole are located in the vicinity of the boundary and additional current is generated.\\

In our example, the function $\mathcal{H}^{K,\sigma\rightarrow K,\sigma}$ has its maximum around $k_\mathrm{out}\ell_{B_0}\sim -1 $.
In general, the position of this maximum determines the contribution of the generated charge carriers to the overall current.
To clarify this statement and its consequence, we consider a general transition with $n_\mathrm{part}=n_\mathrm{in}-1$ and
$\ell_{B_0}k_\mathrm{in}\gg1$ and $n_\mathrm{out}= n_\mathrm{hole}=1$. 
The Coulomb integral (\ref{overlapkkprime}) is exponentially suppressed unless 
$\ell_{B_0}( k_\mathrm{in}- k_\mathrm{part})\ll 1$.
Together with (\ref{largeposk}) and the energy conservation we infer that the energies of the outgoing particle and the generated hole are small,
\begin{multline}
\left(E_{n_\mathrm{out},k_\mathrm{out}}+E_{n_\mathrm{hole},k_\mathrm{hole}}\right)\\
=\frac{v_F}{\ell_{B_0}}\mathcal{O}\left\{
\ell_{B_0}(k_\mathrm{in}
-k_\mathrm{part}),1/\sqrt{\ell_{B_0}k_\mathrm{in}}\right\}\,.
\end{multline}
Thus, the outgoing particle and the generated hole will be 
either zero energy modes of the form (\ref{groundnegk}) or low energy modes 
at the boundary, see equation (\ref{groundsmallk}).

If $n_\mathrm{in}\sim1$ and $\ell_{B_0}k_\mathrm{in}\gg1 $ we know from (\ref{largeposk}) that 
$\Psi_{{k_\mathrm{in}},{n_\mathrm{in}}}^{\mathfrak{p},K}$ and $\Psi_{{k_\mathrm{part}},{n_\mathrm{part}}}^{\mathfrak{p},K}$
are located around the edge and therefore $\mathcal{H}^{K,\sigma\rightarrow K,\sigma} $ will adopt its maximum value at
$|k_\mathrm{out}|\ell_{B_0}\lesssim 1$
if the wavefunctions
$\Psi_{{k_\mathrm{out}},{n_\mathrm{out}}}^{\mathfrak{p},K}$
and $\Psi_{{k_\mathrm{hole}},{n_\mathrm{hole}}}^{\mathfrak{h},K}$
are both dispersive edge modes of the form (\ref{groundsmallk}).
In contrast, for $n_\mathrm{in}\gg 1$ the 
greatest weight of the wave functions $\Psi_{{k_\mathrm{in}},{n_\mathrm{in}}}^{\mathfrak{p},K}$ and $\Psi_{{k_\mathrm{part}},{n_\mathrm{part}}}^{\mathfrak{p},K}$ are close to the classical turning points inside the bulk,
\begin{align}\label{turn}
\frac{|x_\mathrm{turn}|}{\ell_{B_0}}\approx \frac{\sqrt{2 n_\mathrm{in}}}       
{\left(\ell_{B_0}k_\mathrm{in}\right)^{1/4}}\,.
\end{align}
Here $\mathcal{H}^{K,\sigma\rightarrow K,\sigma}$ adopts its maximum value if the outgoing particle and the generated hole are bulk modes.
The position of the maximum of $\mathcal{H}^{K,\sigma\rightarrow K,\sigma}$ can be estimated from (\ref{groundnegk}) and (\ref{turn}) to be at $\ell_{B_0} k_\mathrm{out}\approx -\sqrt{2n_\mathrm{in}}/\left(\ell_{B_0}k_\mathrm{in}\right)^{1/4}$.

In total, we conclude that initial states with small $n_\mathrm{in}$ are beneficial for the current enhancement.
For these dispersive edge modes we have $n_\mathrm{in}/ k_\mathrm{in}\ell_{B_0}\ll 1$ which corresponds to charge excitations traveling nearly parallel to the fold.
In contrast, for large $n_\mathrm{in}$ only the final states which are inside the tail of $\mathcal{H}^{K\rightarrow K}$ at $|\ell_{B_0}k_\mathrm{out}|\lesssim 1$ are relevant for the generated current.

\begin{figure}[t]
\includegraphics[width=\columnwidth]{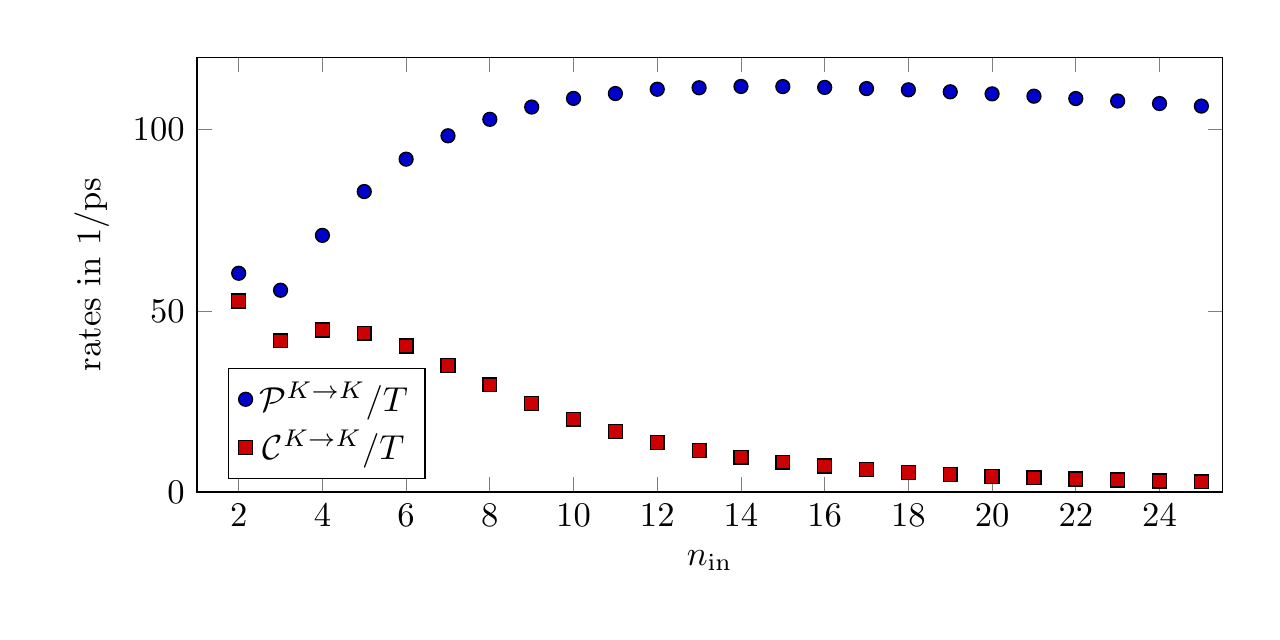} 
\caption{We consider the decay channels from an incoming electron at Landau 
band $n_\mathrm{in}$ with an initial energy $E_{n_\mathrm{in},k_\mathrm{in}}=0.7$ eV
to the outgoing states which are specified by $n_\mathrm{in}\rightarrow n_\mathrm{out}=n_\mathrm{in}-1$ and $ n_\mathrm{part}=n_\mathrm{hole}=1$.
The blue dots denote the generation rates for particle-hole pairs 
%
%
whereas the red squares are the corresponding rates for additionally 
generated charge carriers which contribute to the edge current.}\label{gencurrent}
\end{figure}

In order to quantify our statement, we calculated the 
expectation value of the sum of all currents $J_k=d E/dk$
i.e
\begin{multline}
\frac{\mathcal{C}^{K,\sigma\rightarrow K,\sigma}}{T}=\frac{\alpha^2_\mathrm{graphene}}{8 \pi}
\\
\times \int d k_\mathrm{out} \left(J_{k_\mathrm{out}}+J_{k_\mathrm{part}}+J_{k_\mathrm{hole}}-J_{k_\mathrm{in}}\right)\mathcal{H}^{K,\sigma\rightarrow K,\sigma}
\end{multline}
This quantity can be interpreted as generated particles per unit time which contribute to the edge current.
From Fig.~\ref{gencurrent} we conclude that, although the number of generated charge carriers grows with increasing $n_\mathrm{in}$, the generation of edge modes adopts its maximum value at small values of $n_\mathrm{in}$.
Going back to our initial example we find after summing over all final states
$\mathcal{C}_\mathrm{total}/T\approx 7.8\cdot 10^{14}$s$^{-1}$.
Comparing with (\ref{examplerate}) we conclude that 
only every fourth generated charge carrier will contribute 
to the edge current.
\bigskip
\bigskip
\bigskip
%
%
%
%
%
%

\subsection{Zigzag boundary}

The carrier multiplication process is a rather robust effect and 
we expect that the main characteristics of the decay process also holds 
for a graphene sheet with zigzag boundary.
\begin{figure}[t]
\includegraphics[width=\columnwidth]{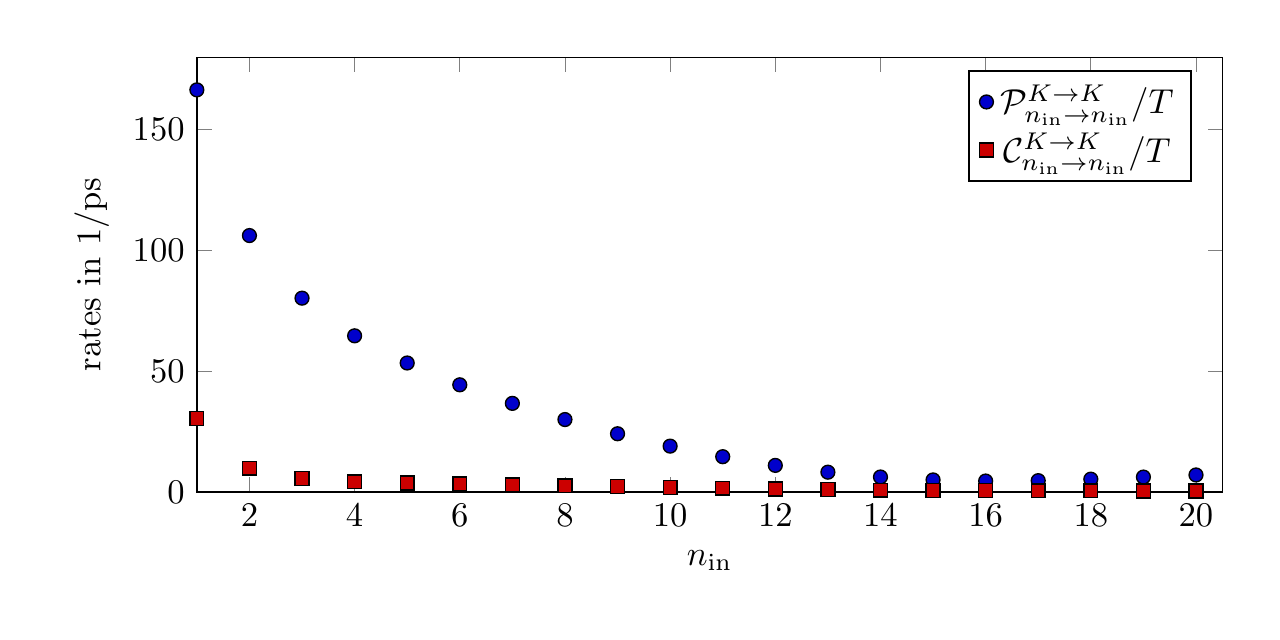} 
\caption{Transitions within the Dirac point $K$ are dominated 
by channels that contain the zero energy mode.}\label{zigzagKKzero}
\end{figure}

\begin{figure}
\includegraphics[width=\columnwidth]{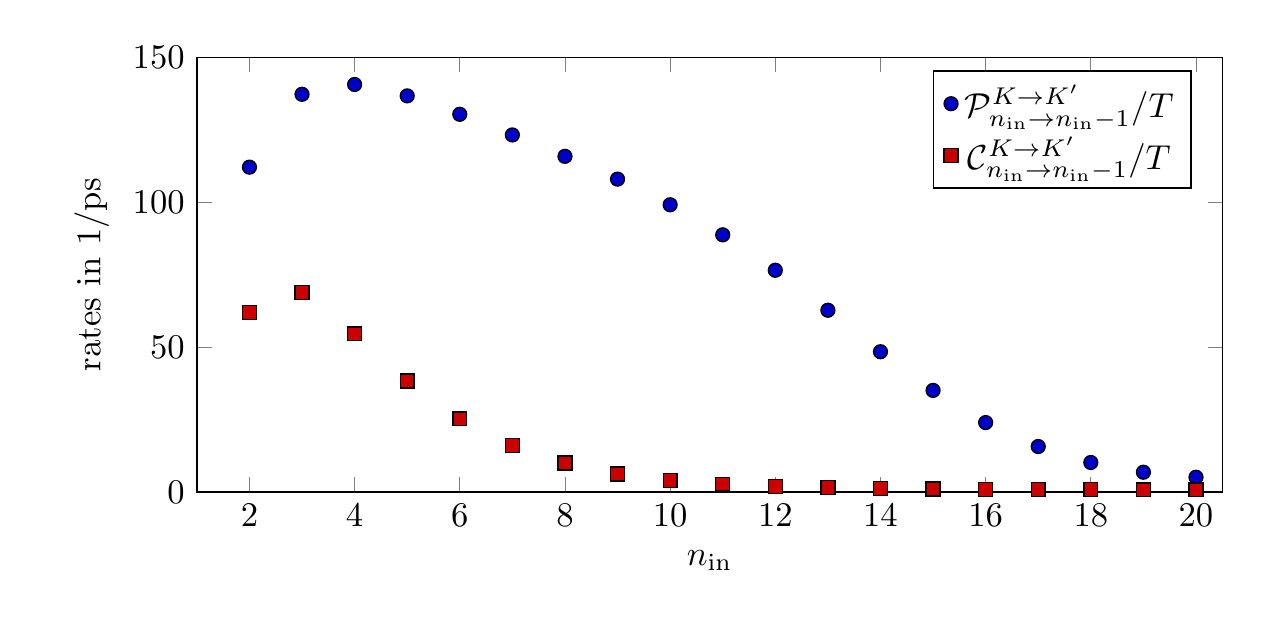} 
\includegraphics[width=\columnwidth]{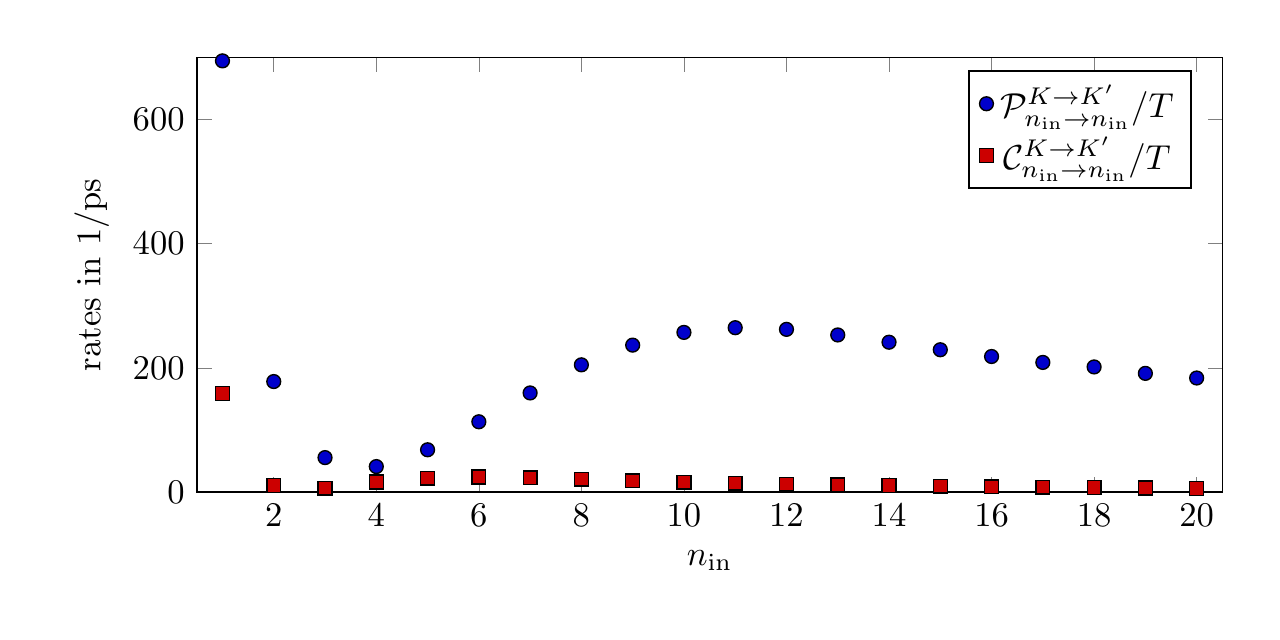}
\caption{Upper panel: Rates for the transitions between the two Dirac points. The quantum numbers for the incoming and outgoing electron are $n_\mathrm{in}$ and $n_\mathrm{out}=n_\mathrm{in}-1$ and the particle-hole pair is generated in the state with $n_\mathrm{part}=n_\mathrm{hole}=0$. The magnitude of these rates are comparable with the results we found in the graphene fold, see Fig.~\ref{gencurrent}.
Lower panel: Rates for channels which are specified by $n_\mathrm{in}=n_\mathrm{out}$ and $n_\mathrm{part}=n_\mathrm{hole}=0$ are dominating the Auger process for small values of $n_\mathrm{in}$.}\label{zigzagKKprime}
\end{figure}

As for the fold geometry, we find that the overlap integrals $\mathcal{I}^{K,\sigma\rightarrow K,\lambda}$
and $\mathcal{I}^{K,\sigma\rightarrow K',\lambda}$ are strongly suppressed unless the approximate selection rule (\ref{selrule}) applies.
Therefore, as for the fold geometry, the charge carrier multiplication is dominated by a few channels.

We assumed for the incoming electron the same energy 
as before, i.e. $E_{n_\mathrm{in},k_\mathrm{in}}=0.7$ eV.
The dominant transitions $K\rightarrow K$ contain 
a zero energy mode $\psi^{K}_{0,k}$ with one vanishing spinor component, see equation (\ref{spinorzigzag}).
The presence of this non-dispersive enlarges the 
phase-space for these channels since momentum-conservation can always be satisfied.
In Fig.~\ref{zigzagKKzero} we show the probabilities per unit time $\mathcal{P}^{K,\sigma\rightarrow K,\sigma'} /T$ for the processes involving the zero energy mode and $n_\mathrm{in}=n_\mathrm{out}$ as well as $n_\mathrm{hole}=1$.
For increasing $n_\mathrm{in}$, the weight of the corresponding wavefunction is moving inside the bulk which
diminishes the overlap with the particle-hole modes at the boundary.
%
%
%
As before, this implies that small values of $n_\mathrm{in}$ are beneficial for the current enhancement.

We found also rather large rates for transitions between both graphene valleys, $K\rightarrow K'$.
Transitions with $n_\mathrm{out}=n_\mathrm{in}-1$ and 
$n_\mathrm{part}=n_\mathrm{hole}=0$ are presented in upper panel of Fig.~\ref{zigzagKKprime}
and are of similar magnitude as the rates in the graphene fold, cf.~Fig.~\ref{gencurrent}.
The largest contribution to the impact ionization 
originates from 
transitions with the quantum numbers $n_\mathrm{in}=n_\mathrm{out}=1$ and 
$n_\mathrm{part}=n_\mathrm{hole}=0$, see the lower panel of Fig.~\ref{zigzagKKprime}.
Note that these transitions are absent in the graphene fold 
due to the pseudo-parity selection rule.

\section{Conclusions and Outlook}

We analyzed the primary magneto-optical absorption
of graphene and the subsequent particle-hole generation
due to impact ionization.
The bare magneto-photoelectric, in particular the 
absorption into the dispersive edge modes, does not exceed 
the well-known value for graphene monolayers.
However, subsequent impact ionization leads
to charge carrier multiplication and therefore 
to a strong enhancement of the photo-current.
We found that charge carrier multiplication 
depends on the incident angle between incoming
electron and graphene edge, particularly 
small impact angles are advantageous 
for current amplification.
The presence of a finite phase space volume due to 
absence of translational invariance makes this effect 
rather robust.
However, the specific enhancement will
depend on the particular boundary condition of the 
graphene edge.

Our findings are consistent with previous investigation 
on the magneto-photoelectric effect \cite{SKG17} and should 
be of relevance for possible future applications such 
as graphene-based photodetectors.

The derived exact and approximate selection rules show that only a small subset of the decay channels will
significantly contribute to the dynamics.
Using this results could permit the efficient implementation 
of the relaxation dynamics using a Boltzmann equation approach.
Although we expect that the qualitative behavior of the charge 
multiplication can already be captured within leading order perturbation 
theory, a quantitative prediction should include higher order 
corrections of the scattering processes.

\acknowledgments 

The authors thank S.~Winnerl and C.~Kohlf\"urst for 
valuable discussions.
Funded by the Deutsche Forschungsgemeinschaft 
(DFG, German Research Foundation) -- Project-ID 398912239.

\end{document}